\documentstyle[12pt,epsfig]{article}          % usual 12pt
\catcode`\@=11
\@addtoreset{equation}{section}

\newcommand{\ccaption}[2]{
  \begin{center}
    \parbox{0.85\textwidth}{
      \caption[#1]{\small\it {#2}}}
  \end{center}    }
        \newdimen\eqskip
        \newdimen\txtskip
        \baselineskip=\txtskip
\def    \be             {\begin{equation}}
\def    \ee             {\end{equation}}
\def    \ba             {\begin{eqnarray}}
\def    \ea             {\end{eqnarray}}

\def    \=              {\;=\;}
\def    \frac           #1#2{{#1 \over #2}}

\def    \ie             {{\em i.e.\/} }
\def    \eg             {{\em e.g.\/} }
\def \as   {\mbox{$\alpha_{\rm S}$}}

\def \ttbar {\mbox{$t \bar t$}}

\def \to   {\mbox{$\rightarrow$}}

\begin{document}
\begin{titlepage}
\nopagebreak
\vspace*{-1in}
{\leftskip 11cm
\normalsize
\noindent
\hfill hep-ph/9604332 \\
}
\vskip 1.5cm
\begin{center}
{\LARGE  \bf \sc QCD in $e^+e^-$ Collisions at 2~TeV}
\\[3cm]
{\bf Sigfried Bethke$^a$,
Michelangelo L. Mangano$^b$\footnotemark and Paolo Nason$^b$\footnotemark}
\\[2cm]
\footnotetext{On leave of absence from INFN, Sezione di Pisa, Pisa, Italy.}
\footnotetext{On leave of absence from INFN, Sezione di Milano, Milano, Italy.}
$^a$ III. Phys. Inst, RWTH,  D-52074 Aachen, Germany \\
$^b$ CERN TH-Division, CH-1211 Geneva 23, Switzerland \\
\vskip 0.3cm
\end{center}
\vskip 1cm
\nopagebreak
\vfill
\begin{abstract}
{\small
We discuss some topics in QCD studies at 2~TeV. Particular emphasis is
given to the separation of pure QCD events from the $WW$ and the
\ttbar\ backgrounds.
}
\end{abstract}
\vfill
\noindent
%CERN-TH.XXXX/95 \newline
%October 1995    \hfill
\end{titlepage}
\section{Introduction}
QCD studies in $e^+e^-$ colliders at 2 TeV will be similar in many
respects to QCD studies at lower energy (for a review of studies in
$e^+e^-$ collisions at 500~GeV, see \eg\ ref.~\cite{previous}).  One
will attempt to measure the strong coupling constant, scaling
violation in fragmentation functions, detailed studies of multi-jet
distributions, and the like.  In this short study we concentrate of
features which are novel to collisions at this high energy. The
general theoretical and experimental framework for QCD studies in
$e^+e^-$ collisions can be found in previous reviews \cite{previous}.

The main problem at this energy, is how to disantangle the
$t\bar{t}$ and $W^+W^-$ backgrounds from the pure QCD di-jet or
multi-jet events. The $W^+W^-$ cross section is
particularly high and when the $W$'s decay into hadrons
one will be unable to distinguish the final state from that of two
ordinary high energy jets. A typical 1 TeV QCD jet will in fact have
an invariant mass of the order of $\as$ times 1 TeV, which is roughly
90 GeV, of the same order of the $W$ mass.
\begin{figure}[htb]
\centerline{\psfig{figure=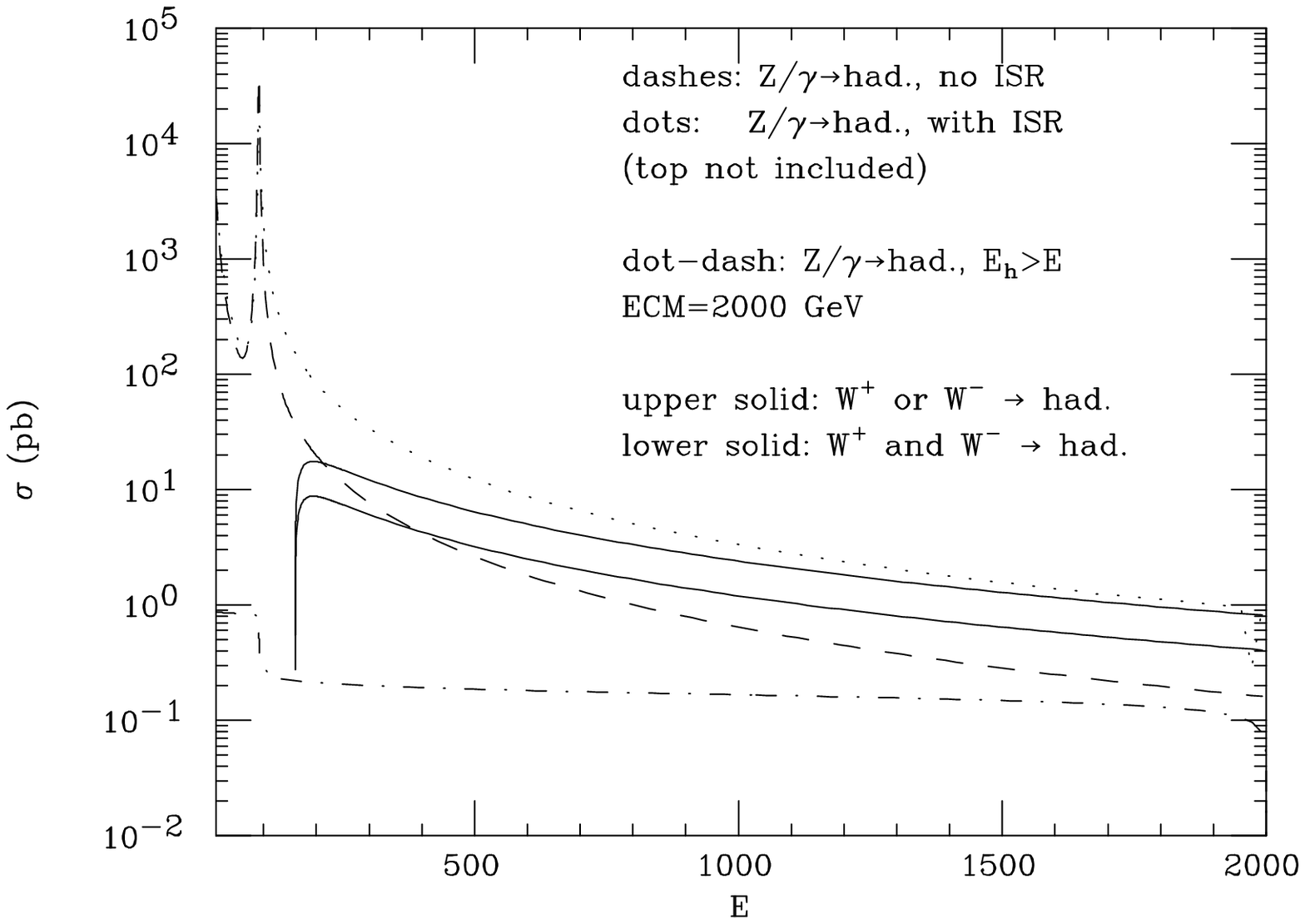,width=0.9\textwidth,clip=}}
\ccaption{}{ \label{f1}
Cross sections for $e^+e^-$ into light hadrons and into $WW$ pairs, as
a function of
$\sqrt{s}$. The solid lines represent the $WW$ cross section
(upper curve: at least one $W$ hadronic decay; lower curve: both $W$
hadronic decays). The light hadrons cross section
is given by the dotted line (no initial state radiation), and by
the dashed line (ISR included). The dash-dotted line is the integral
of the light hadron cross section at $\sqrt{s}=2$~TeV, integrated
above a given value of the energy of the final hadronic state, in
presence of ISR.}
\end{figure}
In fig.~\ref{f1} we show
the cross section for $e^+e^-$ into light hadrons (from now on, by
``hadronic events'' we will mean $e^+e^-$ into light hadrons, \ie\
top contribution excluded), and the $WW$ cross section. 
Initial state radiation causes a loss of cross section around
a factor of two, depending on the cuts one imposes. In the figure the
ISR effect is only shown for hadronic events, but we should expect
something similar for the $W$. It is therefore reasonable to assume
that the Born cross sections give a good indication of the relative
magnitudes of the $WW$ and hadronic cross sections. We should
therefore compare 0.16~pb for the hadronic cross section to 0.4~pb for
the $WW$ fully hadronic events. Fortunately, the $WW$ events are
concentrated in the forward region, because of their $t$-channel nature.
\begin{figure}[htb]
  \centerline{\psfig{figure=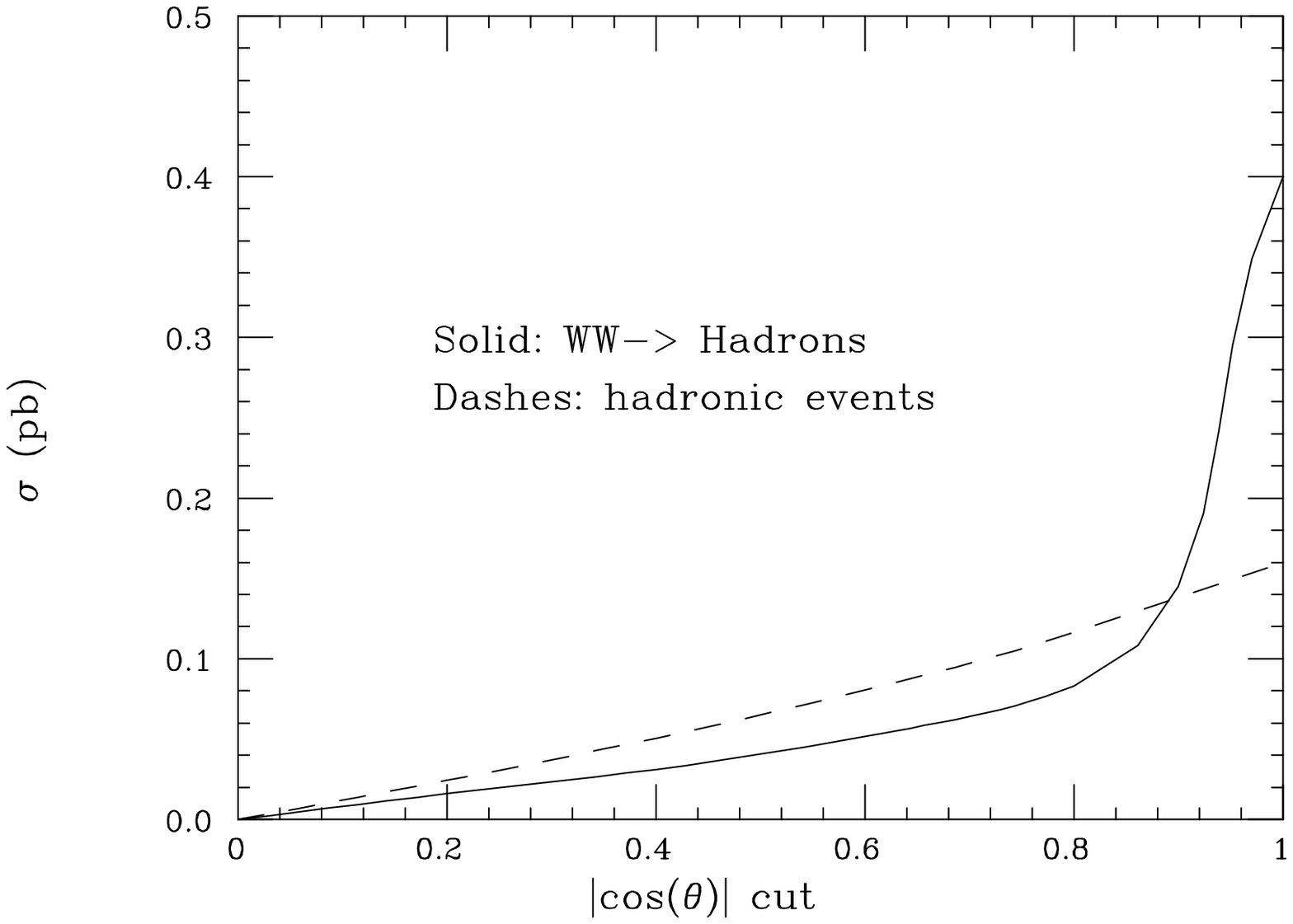,width=0.9\textwidth,clip=}}
  \ccaption{}{ \label{f2} Cross section at 2~TeV as a function of the
    cut on the direction of the thrust axis.}
\end{figure}
This is shown in fig.~\ref{f2}, where the cross section is shown as a
function of an angular cut on the thrust axis. We see that requiring
$\cos\theta<0.8$ leaves us with a 60\% of the hadronic events and 40\%
of $WW$ events. Further work is needed to clean the sample in a better
way. For example, one expects in general that $WW$ events will always
have high thrust. Conversely, one can assume that the $WW$ events are
well understood ($W$ decays will be similar to $Z$ decays, a fact that
can be verified at LEP2), and subtract them from the measured
distributions.

\begin{figure}[htb]
  \centerline{\psfig{figure=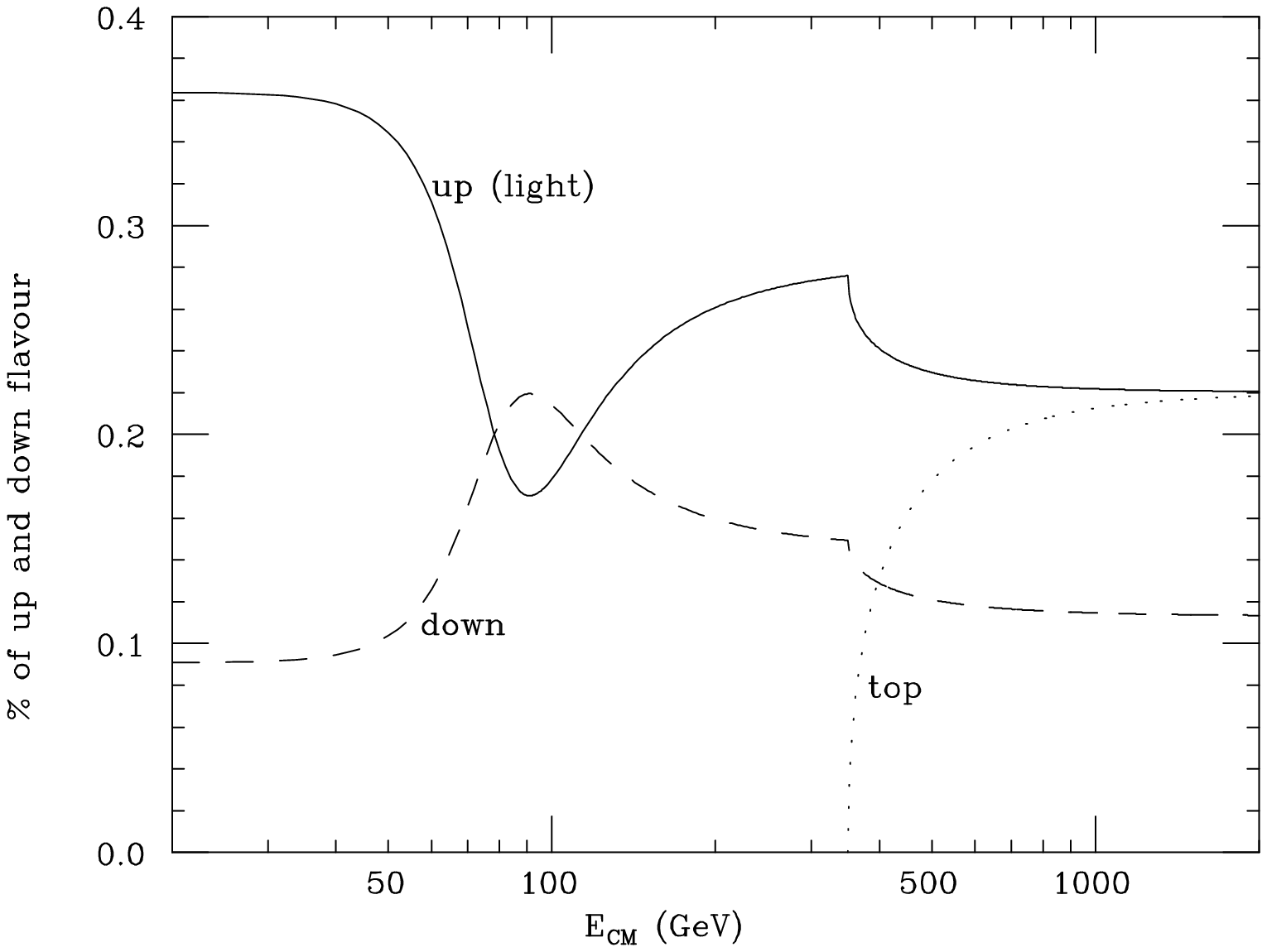,width=0.9\textwidth,clip=}}
  \ccaption{}{ \label{f3} Flavour composition for $Z/\gamma \to q \bar
    q$ as a function of $\sqrt{s}$.}
\end{figure}
The separation of \ttbar\ events will also require some work. The
relative fraction of each flavour is given in fig.~\ref{f3}. At 2~TeV
the top quark production represents 22\% of the total.

In the following sections we will study some simple variables which
allow to separate the various contributions, and will discuss to which
extent a separation is indeed necessary. A complete study of the
potential for QCD studies at 2~TeV would require some detailed
knowledge of the detector specifications. In what follows, we will
therefore briefly comment on which detector parameters have an impact
on this physics.

\section{Conventions}
We will define jets using the Durham algorithm \cite{durham},  in the E
recombination scheme. Unless otherwise stated, the MC studies are done using
HERWIG \cite{herwig}  version 5.8, and use partons as opposed to hadrons.
Namely, jets are reconstructed out of the partons after the full shower
evolution and gluon splitting, and before being clustered and hadronized.

No QED ISR nor beamstrahlung have been included, as our assigned  goal was to
study QCD at {\em exactly} (or very close to) 2~TeV.

A word of caution must be added, however, before we present the results of our
case study: here, we are extrapolating our current knowledge to c.m. energies
which are about {\it twenty times larger} than those energies which are presently
available.
This is equivalent to a hypthetical attempt to
predict the physics at LEP-I based on the knowledge of the early data from
SPEAR ($\sqrt{s} \sim 5$~GeV).

\section{Separation of QCD and $WW$ events}
It is hard to separate the two samples without significantly biasing the
properties of the jets. The typical property of a $WW$ event is that no
hard emission at large angle is possible. In fact the $W$'s will decay in
flight into $q\bar q'$ pairs, and the mass of the hadronic system resulting
will never exceed $M_W$. Since the $W$'s are boosted to 1 TeV, the two or more
jets from their decay will coalesce into a single thin jet, with angular
aperture of the order of $M_W/1\;{\rm TeV} \sim 5^\circ$. Particles emitted
outside this cone cannot be too energetic, or else they would form together
with the leading jet an object of invariant mass higher than $M_W$. We tried to
use this property to separate $WW$ events from QCD events.
\begin{figure}[htb]
\centerline{\psfig{figure=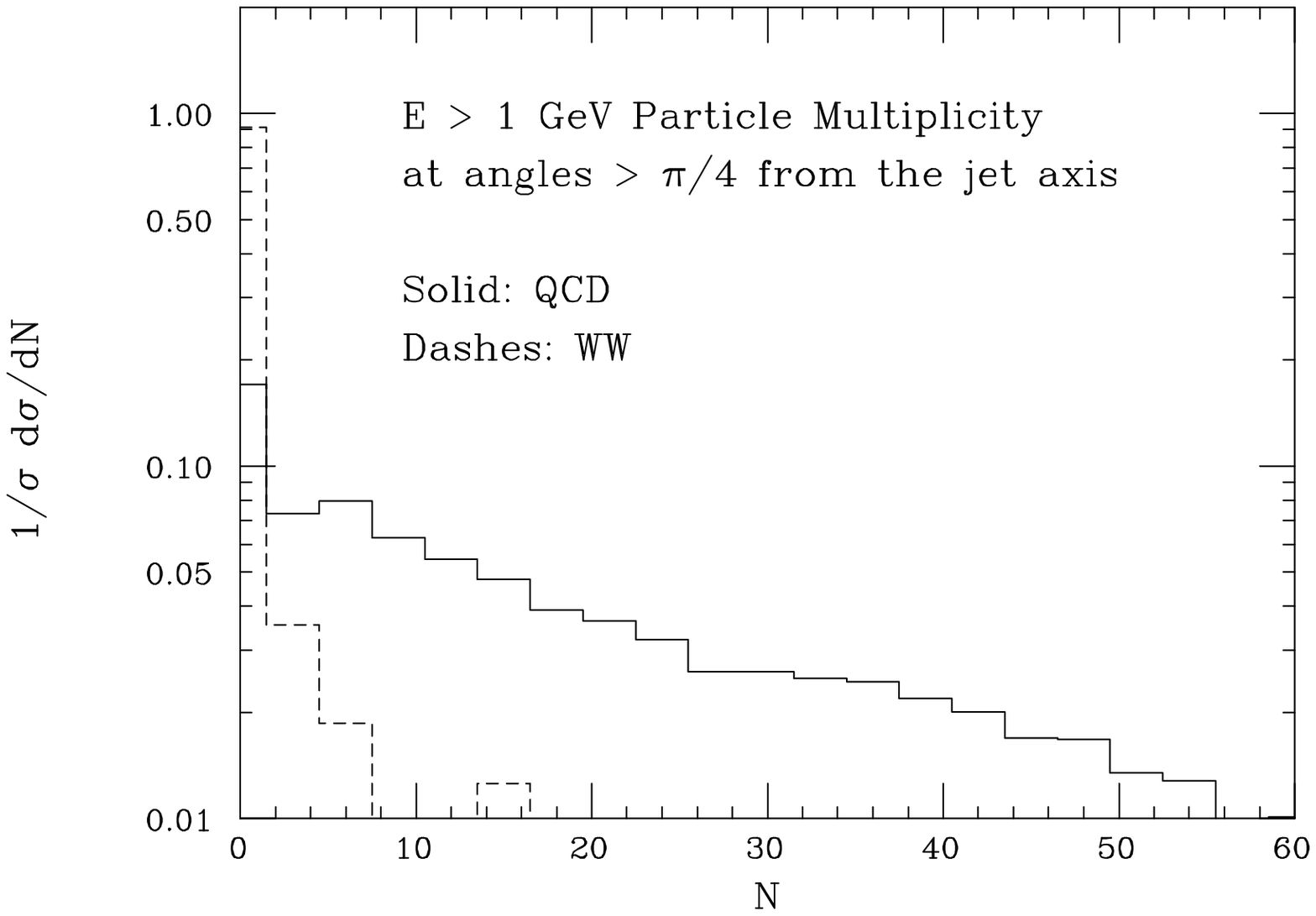,width=0.4\textwidth,clip=},
            \psfig{figure=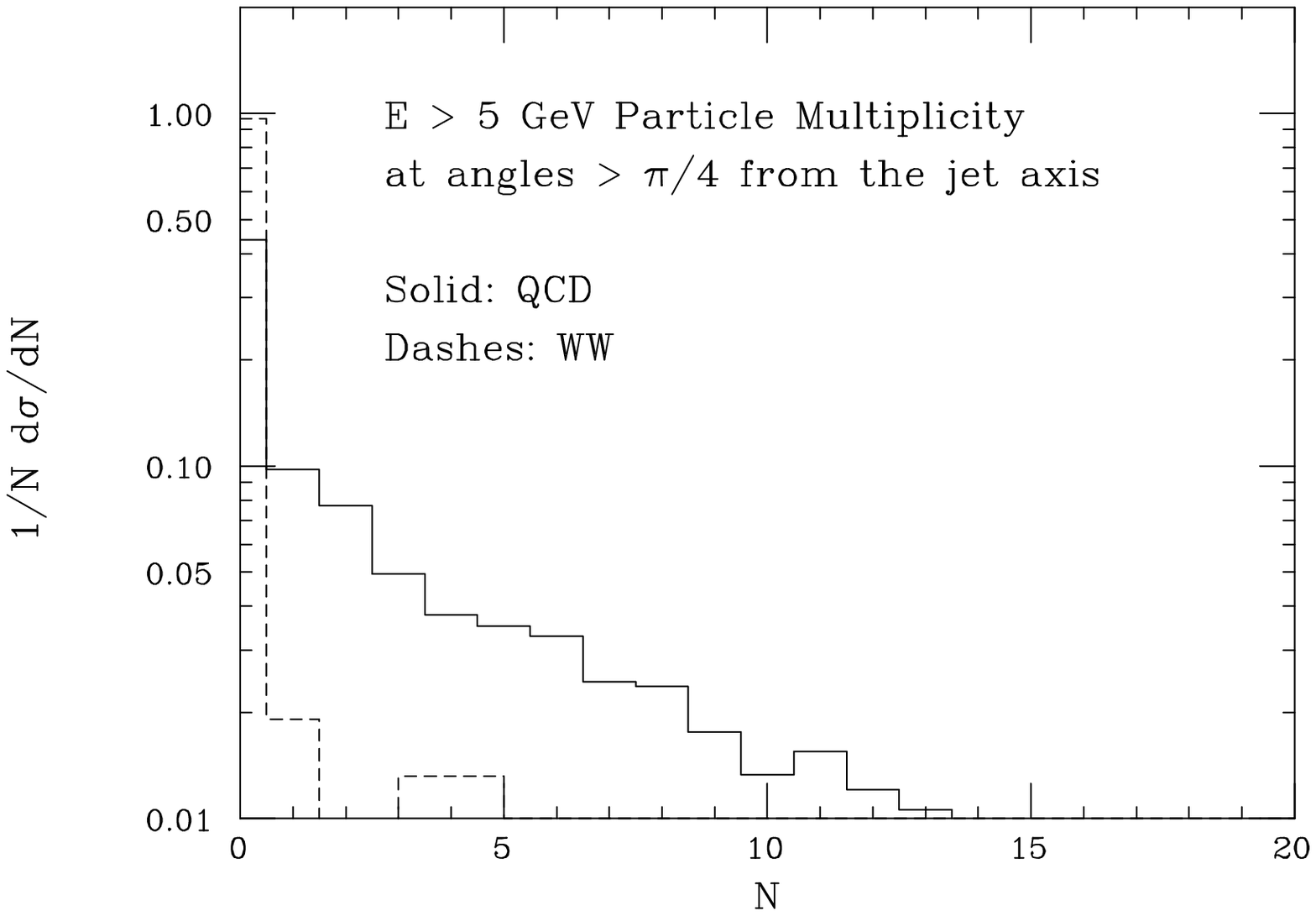,width=0.4\textwidth,clip=}}
\ccaption{}{ \label{f-Wcone}
Multiplicity distribution for particles above 1~GeV (left) and 5~GeV (right)
found in the region outside cones of radius $45^\circ$ centered around the
axis of the two leading jets. QCD events (solid) vs. $WW$ events (dashed).
}
\end{figure}
In fig.~\ref{f-Wcone}\ we plot the multiplicity distribution of
particles above 1 and 5~GeV found in the region outside cones of
radius $45^\circ$ centered around the axis of the two leading jets.
The continuous lines are for QCD events, the dashed lines for $WW$
events. From the figures we learn the following:
\begin{itemize}
\item The area outside the jet cores is indeed much quieter in $WW$ events.
\item Only a fraction of the order of 10\% of the $WW$ events would survive the
request that particles above 1 GeV be present outside the $45^\circ$ cones.
More than 80\% of the QCD sample would survive this cut.
\item If we require the 5 GeV cut, only a fraction of the order of $10^{-2}$ of
the $WW$ events would survive. Approximately 50\% of the QCD sample would
be left.
\item It is not clear how this request would bias the
jets, and whether the extraction of \as\ from the properties of jets selected
in this way would have a significant systematic uncertainty.
This issue can however be
studied using shower Montecarlo programs.
\end{itemize}

\section{Jet Production Rates and $t\overline{t}$ Events}
For the studies presented in this section we generated hadronic final states
at $\sqrt{s} = 2$~TeV
using PYTHIA version 5.7 \cite{jetset} with QCD- and hadronisation parameters as
optimised to describe the data from LEP1.
As before, effects due to initial state radiation, beamstrahlung and detector
acceptance are not taken into account.
\begin{figure}[htb]
\centerline{\psfig{figure=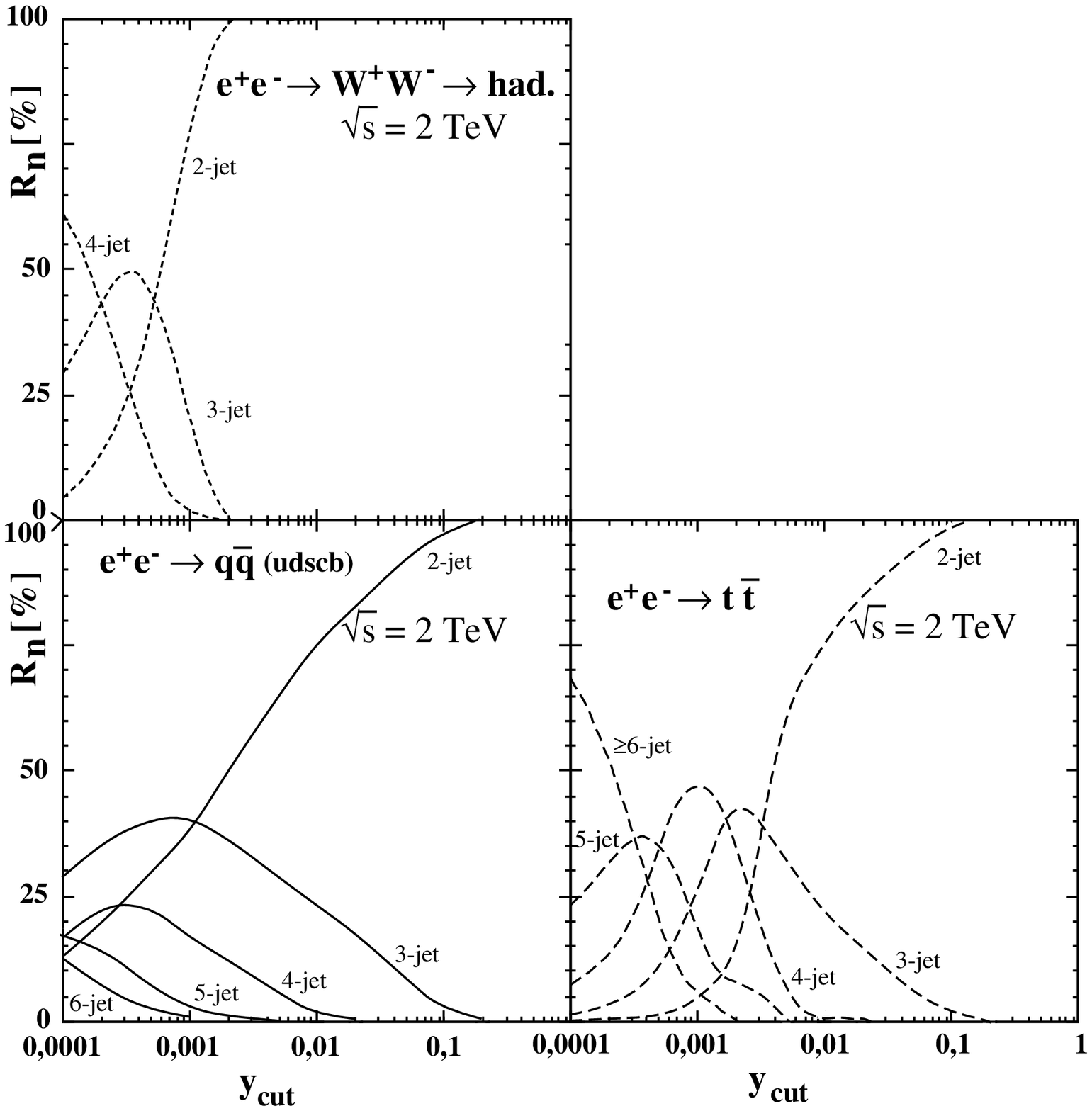,width=14cm,clip=}}
\ccaption{}{ \label{f-siggi}                
Integral relative production rates of n-jet events (n~=~2, 3, 4, 5,
$\ge$6), as a function of the jet resolution parameter $y_{cut}$.}
\end{figure}
The Durham ($k_{\perp}$) jet algorithm \cite{durham} is used to study jet
multiplicities for different event classes, namely for hadronically decaying
W-pair events ($e^+e^- \rightarrow W^+W^- \rightarrow $~hadrons), for hadronic
events from primary `light' quark pairs ($e^+e^- \rightarrow q\overline{q}$;
where $q$ may be a $u$, $d$, $s$, $c$ or $b$-quark) and for top quark events
($e^+e^- \rightarrow t\overline{t}$).
The results are shown in Fig.~\ref{f-siggi}, where the
integral, relative production rates of n-jet events (n~=~2, 3, 4, 5,
$\ge$6) are plotted as a function of the jet resolution parameter $y_{cut}$.
The following observations can be made:

\begin{itemize}
\item for $y_{cut} > 0.002$, all $W^+W^-$ events are classified as 2-jet
($\sqrt{0.002} \times 2000$~GeV = 98.4~GeV $> M_W$).
\item $t\overline{t}$-events have a markedly enhanced and clear 6-jet signal
around $y_{cut} \sim 0.0001$. This allows to select top-quark events and study
e.g. the decay $t \rightarrow $~3~jets in detail. Of course, the
reconstruction of jets at such small values of $y_{cut}$ might require
particular granularity in the calorimeter. The accuracy with which
the top mass will be measured from the reconstruction of these high
energy jets could significantly depend on this parameter.
\item for $y_{cut} > 0.01$, the 3-jet production rates of $t \overline{t}$ and
$q \overline{q}$ events are almost identical - which is intuitively clear
since $\sqrt{0.01} \times 2000$~GeV~$\ge M_{t}$.
\end{itemize}

\noindent From these observations and other studies we conclude:

\begin{itemize}
\item At c.m. energies of 2000~GeV,
the separation of $t\overline{t}$ events from light-quark QCD events for
a typical QCD analysis, like e.g. the determination of $\as$ from 3-jet
rates, seems not to be necessary: these event
classes show similar QCD properties in regions where the jet resolution is
coarser than the mass of the top quark. The situation is similar as for b-quark
events at e.g. TRISTAN energies ($\sqrt{s} \le 60$~GeV), where the mass of the
heaviest quark also was about 10\% of the c.m. energy.
\item In fact we tried to separate $t \overline{t}$ events from light quark
QCD events using kinematic variables, similarly to  the
separation of $W^+W^-$ events described in the previous section.
While the algorithms which we devised are good in extracting a sample of top
events with good purity, they are not very good at providing a pure sample of
light quark QCD events which is not too biased. The same is true when
selecting $t\overline{t}$ events by requiring 6 jets at $y_{cut} = 0.0001$; see
Fig.~\ref{f-siggi}.
\item Determination of $\as$ from 3-jet event rates will be possible by
analysing {\em relative} production rates or by analysing {\em absolute}
cross sections of 3-jet events. In the first case, one has to correct the
measurement according to the large production of $W^+W^-$ events, which all end
up as 2-jets for decent values of $y_{cut}$ and therefore spoil the relative
number of 3-jets by the overall normalisation. The absolute normalization of
the $WW$ rate is well known within the SM, this being a purely EW process. If
we allow for possible new phenomena, we could still determine the $WW$ rate
from the data, by counting the number of events where one of the $W$'s decays
leptonically. The statistical error of this measurement is comparable to the
statistical error on the QCD 2-jet production rate.
The use of the {\em absolute} 3-jet event production rate is not influenced by
the production of 
$W^+W^-$ pairs, provided one works with $y_{cut}>0.002$.  This method however
relies on a good knowledge of luminosities, acceptances and theoretical
expectations. The uncertainty on these last ones is in principle
correlated to that at LEP1 and LEP2, and should not constitute a major source
of systematics for comparisons between results at LEP1/LEP2 and NLC.
\item With 5000 selected hadronic ($q\overline{q}$ and $t\overline{t}$) events
(i.e. after about 2-3 years of data taking with luminosities around $10^{33}
s^{-1} cm^{-2}$), we expect about 750 3-jet events at $y_{cut} = 0.02$, leading
to a statistical precision in $\as$ of about 3\%. Including the systematic
sources mentioned in the previous point, we would estimate the final
uncertainty on $\as$ not to exceed 5\%.
Starting with $\as(M_Z) = 0.123\pm 0.002$ we evaluate an expected
$\as(2~{\rm TeV}) = 0.085\pm 0.004$. The evidence for running would
therefore be a clear 7~$\sigma$ effect. On the contrary,
with the value preferred by DIS ($\as(M_Z)  = 0.111$) as a boundary
condition one should expect  $\as(2~{\rm TeV}) = 0.078\pm 0.004$. The
difference between the two values at 2~TeV
is less than 2~$\sigma$. As
expected,  operations at 2~TeV will not enable any improvement in the
measurement of $\Lambda_{QCD}$.
\end{itemize}

\section{Fragmentation Functions}
\begin{figure}[htb]
\centerline{\psfig{figure=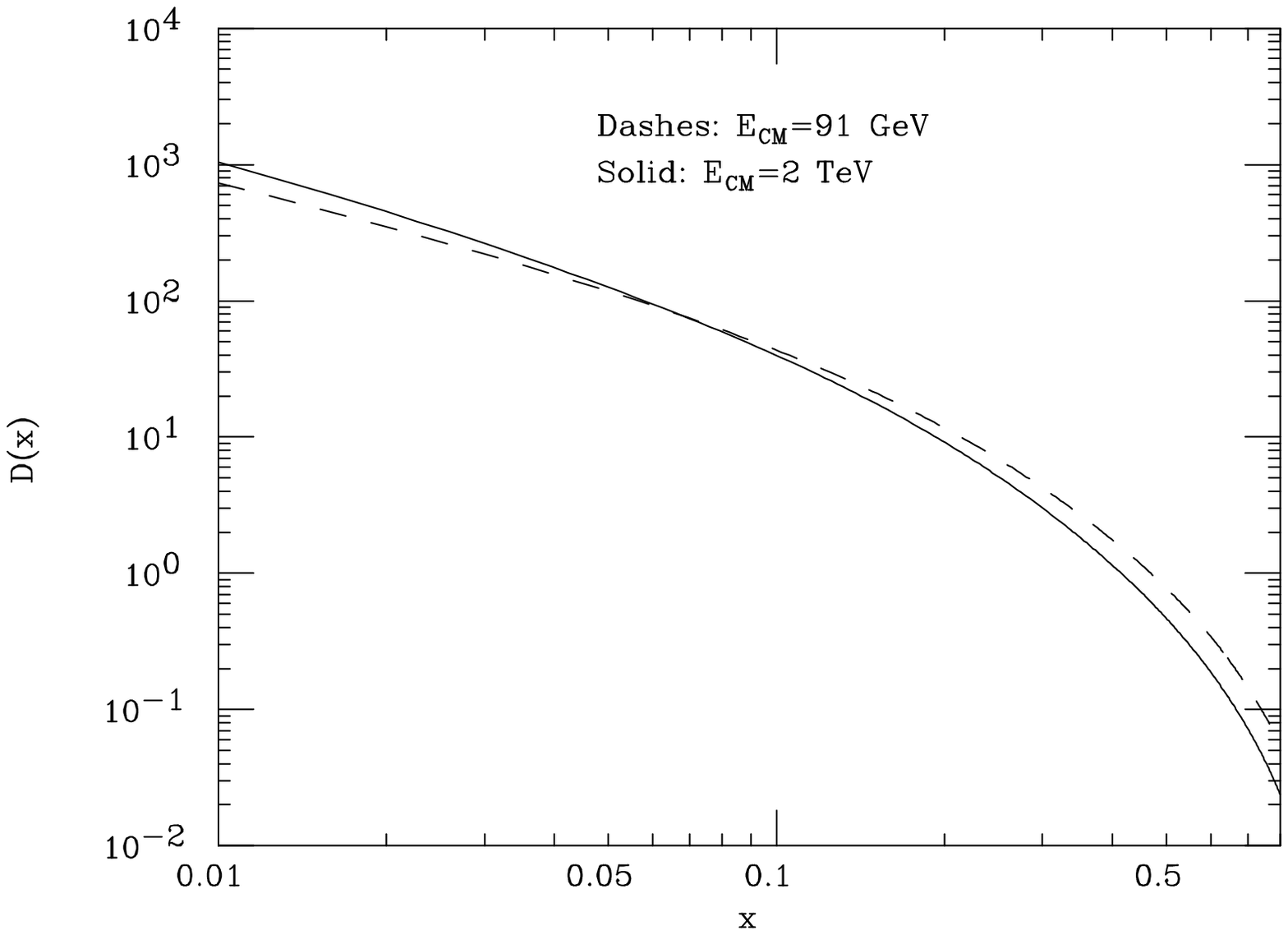,width=14cm,clip=}}
\ccaption{}{ \label{f-frag}
Inclusive charged particle fragmentation function at $\sqrt{s}=M(Z)$ (dashed)
and 2~TeV (solid).}
\end{figure}
The study of the evolution of fragmentation functions provides another
interesting test of perturbative QCD. In fig.~\ref{f-frag}\ we show the
expected evolution of the inclusive fragmentation functions from 
$\sqrt{s}=M_Z$
to 2~TeV. Although the differences are quite noticeable (the fragmentation
function is softer by almost a factor of 2
at $z>0.5$ because of the larger amount of radiation given off), it is not
clear to which extent such effects can be measured. The highest energy jets
will be extremely collimated, and fundamental parameters such as track
reconstruction efficiency, fake track rates and momentum resolution will
depend very strongly on the detector parameters: magnetic field, tracking
resolution etc. It is therefore impossible at this stage to formulate a
potential for a physics measurement based on fragmentation functions.

\begin{figure}[htb]
\centerline{\psfig{figure=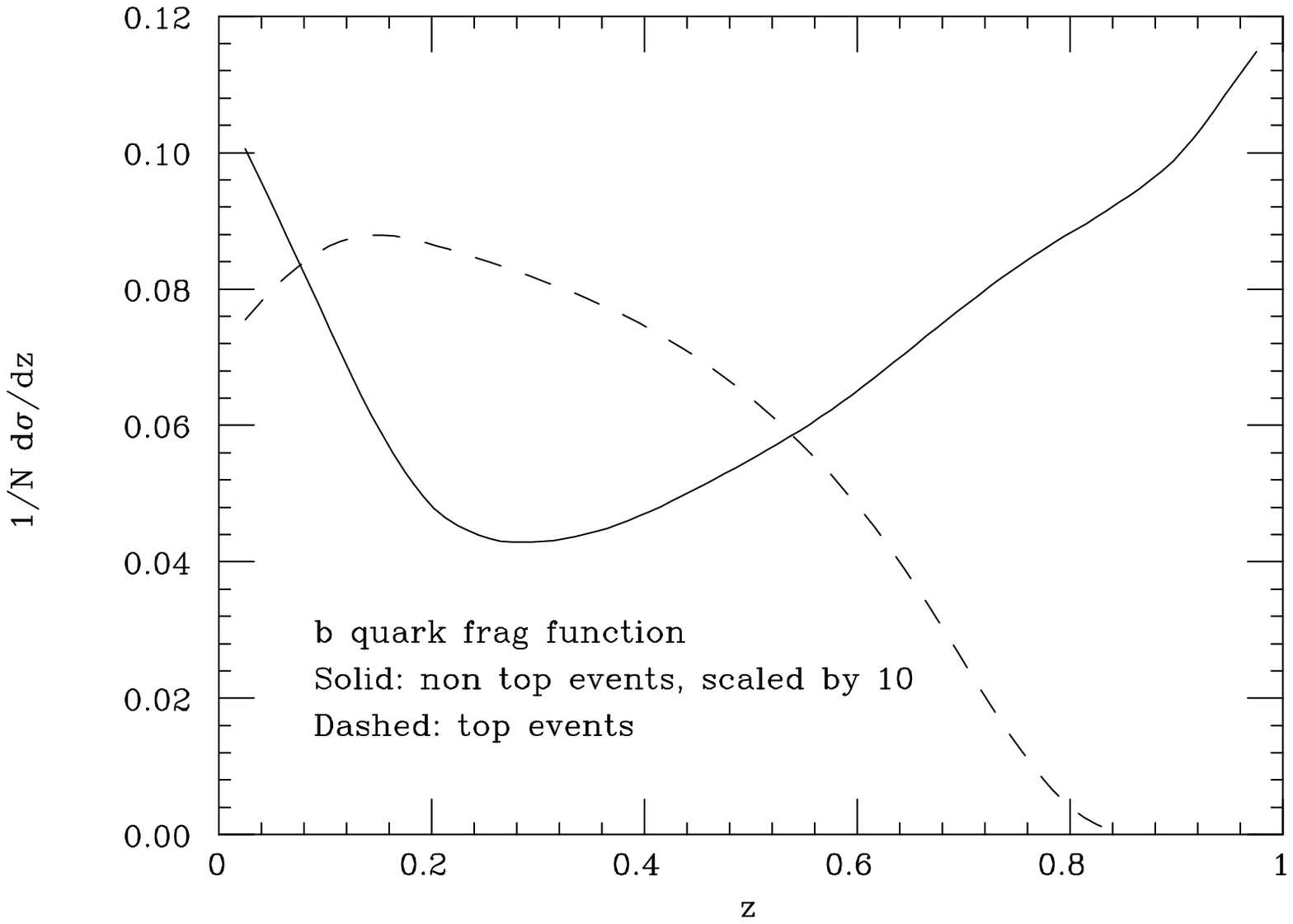,width=14cm,clip=}}
\ccaption{}{ \label{f-bfrag}
Inclusive $b$ quark fragmentation function at 2~TeV.
Non-$t\bar t$ events (solid) and $t\bar t$ events scaled by 10 (dashed).}
\end{figure}

In fig.~\ref{f-bfrag}\ we plot the $b$ fragmentation function, separated
between non-$t\bar t$ events and  $t\bar t$ events.
We point out the following features:
\begin{itemize}
\item The high tail at small-$z$ in non-\ttbar\ events  comes from the
splitting of gluons emitted during the evolution of light $q\bar q$ events.
\item A large fraction of inclusive $b$'s therefore  comes from non-$b$
events. Measurements of the direct $Zb\bar b$ vertex at 2~TeV will therefore be
severely biased by our capability to predict the precise contamination due to
events initiated by light quarks. The requirement of double tagging on opposite
jets would significantly reduce the non-$Zb\bar b$ signal, but at the expense of
a loss in statistics.
\item In addition, the only region of $z$ in which a pure sample of non-\ttbar\
events can be selected is for $z>0.8$. Below this value, the two fragmentation
functions do not differ enough to allow an event-by-event separation of the two
components.
\item The average momentum of a $b$ is of the order of $z=1/2$, i.e. 500 GeV,
both in \ttbar\ and in non-\ttbar\ events. This is more than
ten times the momentum of $b$'s produced at LEP1, and corresponds to decay
lenghts of the order of 4~cm.
The secondary-vertex tagging detectors should therefore be optimized
accordingly. In particular, the radius should be such as to
guarantee that the acceptance for the decay taking place {\em before}
the tracking device be large enough. Furthermore, in comparison with LEP1
$b$'s at 2~TeV will be softer relative to the remaining tracks of the jet, and
surrounded by larger hadronic activity.
\end{itemize}
Altogether, it is therefore very difficult to estimate the impact
of the detector design on the physics potential.  Simple extrapolations from
LEP1 values could be seriously misleading, and realistic designs and tracking
reconstruction algorithms will have to be used before numbers such as
$b$-tagging efficiency can be estimated.


\begin{thebibliography}{99}
\def    \nuke   #1#2#3{{\sl Nucl. Phys.} {\bf B#1}  (#2) #3}
\def    \pl     #1#2#3{{\sl Phys. Lett.} {\bf #1B}  (#2) #3}
\def    \prl    #1#2#3{{\sl Phys. Rev. Lett.} {\bf #1}  (#2) #3}
\def    \pr     #1#2#3{{\sl Phys. Rev.} {\bf #1}  (#2) #3}
\def    \prd    #1#2#3{{\sl Phys. Rev.} {\bf D#1}  (#2) #3}
\def    \prep   #1#2#3{{\sl Phys. Rep.} {\bf #1}  (#2) #3}
\bibitem{previous}
S~Bethke et al., in ``$e^+e^-$ Collisions at 500 GeV: the Physics Potential'',
P. Zerwas ed., DESY 92-123A, Vol.~A, p.393-463;\\
G.~Cowan, in ``$e^+e^-$ Collisions at 500 GeV: the Physics Potential'',
P. Zerwas ed., DESY 93-123C, Vol.~C, p.331.
\bibitem{durham}
  S. Catani et al., \pl{269}{1991}{432}.
\bibitem{herwig}
        G. Marchesini and B.R. Webber, \nuke{310}{1988}{461}.
\bibitem{jetset}
        T. Sj\"ostrand, {\it Comput. Phys. Commun.} {\bf 82} (1994) 74.
\end{thebibliography}
\end{document}